\documentclass[aps,prl,twocolumn,superscriptaddress,amsmath,nofootinbib,natbib]{revtex4-1}

\usepackage[hidelinks]{hyperref}
\usepackage{amsfonts}
\usepackage{mathtools}
\usepackage{aas_macros}
\usepackage[capitalize]{cleveref}                                                             
\usepackage[export]{adjustbox}
\usepackage{lipsum}

\Crefname{equation}{Equation}{Equations}
\crefname{equation}{Eq.}{Eqs.}
\Crefname{figure}{Figure}{Figures}
\crefname{figure}{Fig.}{Figs.}
\Crefname{table}{Table}{Tables}
\crefname{table}{Tab.}{Tabs.}
\Crefname{section}{Section}{Sections}
\crefname{section}{Sec.}{Secs.}

\usepackage{xcolor}

\renewcommand{\d}[2][]{\operatorname{d}^{#1}\!{#2}}

\begin{document}

\title{Quantifying the global parameter tensions between ACT, SPT and \textit{Planck}}

\author{Will Handley}
\email[]{wh260@mrao.cam.ac.uk}
\affiliation{Astrophysics Group, Cavendish Laboratory, J.J.Thomson Avenue, Cambridge, CB3 0HE, UK}
\affiliation{Kavli Institute for Cosmology, Madingley Road, Cambridge, CB3 0HA, UK}
\affiliation{Gonville \& Caius College, Trinity Street, Cambridge, CB2 1TA, UK}
\homepage[]{https://www.kicc.cam.ac.uk/directory/wh260}

\author{Pablo Lemos}
\email[]{pablo.lemos.18@ucl.ac.uk}
\affiliation{Department of Physics and Astronomy, University College London, Gower Street, London, WC1E 6BT, UK}

\newcommand{\WH}[1]{WH: \textbf{#1}}
\newcommand{\PL}[1]{PL: \textbf{#1}}

\newcommand{\WHTODO}[1]{\textbf{WH TODO: #1}}
\newcommand{\PLTODO}[1]{\textbf{PL TODO: #1}}
\newcommand{\TODO}[1]{\textbf{TODO: #1}}
\date{Submitted 16\textsuperscript{th} July 2020}

\begin{abstract}
    The overall cosmological parameter tension between the Atacama Cosmology Telescope 2020 fourth data release (ACT) and \textit{Planck} 2018 data 
within the concordance cosmological model is quantified using the Suspiciousness statistic to be $2.6\sigma$. Between ACT and the South Pole Telescope (SPT) we find a tension of $2.4\sigma$, and $2.8\sigma$ between ACT and Planck+SPT combined. 
While it is unclear whether the tension is caused by statistical fluctuations, systematic effects or new physics, caution should be exercised in combining these cosmic microwave background datasets in the context of the $\Lambda$CDM standard model of the universe.
\end{abstract}

\pacs{}
\maketitle
\section{Introduction}\label{sec:introduction}
As cosmological datasets increase in quantity and quality, so does our capacity to use them to pin down the properties of our universe~\cite{Scott:2018adl}. The error bars on the measurements of cosmological parameters have narrowed over recent years and discrepancies between datasets (or ``tensions'') have begun to emerge. Whilst this is most stark when examining differing observations of the Hubble parameter between early and late time cosmological probes~\cite{2014MNRAS.440.1138E, 2018JCAP...09..025M, 2019ApJ...876...85R}, other more minor tensions arguably exist in clustering parameters between weak lensing and the cosmic microwave background (CMB)~\cite{2019arXiv190105289D, 2020A&A...633A..69H} and in cosmic curvature between the CMB and CMB lensing/Baryon Acoustic Oscillations~\cite{2019arXiv190809139H,2020NatAs...4..196D, 2020MNRAS.496L..91E}. 

When a substantial tension occurs, it may indicate either a systematic error in how either or both of the datasets have been gathered and analysed, or more excitingly may hint at evidence for new physics if extensions or modifications to our concordance model can bring the inferred parameters back into alignment.

In the case of the ``Hubble tension'' where a single obvious cosmological parameter such as the present day expansion rate $H_0$ is discrepant by $\sim5\sigma$, there is little doubt that something is fundamentally wrong. The other tensions are more subtle, in that they are only visible in complicated combinations of the parameters. As shown by \cref{fig:parameters}, in modern cosmology, error bars on the parameters of our universe are represented by high-dimensional Bayesian probability distributions. Visualising a ``distance'' between these degrees of belief is challenging, and in recent years a good deal of theory has been developed for defining a variety of metrics of discrepancy~\cite{2017PhRvD..95l3535C, 2019arXiv190910991L}.

The latest Atacama Cosmology Telescope (ACT) data release 4~\cite{2020JCAP...12..047A,2020JCAP...12..045C,2020JCAP...12..046N} represents the most recently acquired CMB data, with two other measurements of the CMB power spectrum across a wide range of multipoles being provided by the \textit{Planck} satellite~\cite{2020A&A...641A...5P,2020A&A...641A...6P}, and the South Pole Telescope (SPT)~\cite{2018ApJ...852...97H}. By eye it is clear that in the ACT data some parameters such as the spectral tilt of the primordial power spectrum $n_s$ are mildly discrepant, but it is always possible in a high dimensional parameter space that such discrepancies occur by chance and are unremarkable. 

In this letter we discuss how this tension is quantified rigorously using the global Suspiciousness statistic~\citep[][henceforth H19]{2019PhRvD.100d3504H}, and find that ACT is in mild-to-moderate tension with \textit{Planck} and SPT, at a similar or greater level to that found in weak lensing data. We place ACT's own global tension analysis in the context of the tensions literature, and extend it by considering SPT data and further emphasise the perils of focussing too closely on lower-dimensional views onto the cosmological constraints.

\section{Methodology}\label{sec:methodology}


Quantifying tension between high dimensional posterior distributions is a non-trivial problem, even under the approximation of a Gaussian distribution. This has led to a large number of papers describing methods to quantify tension in high dimensional problems~\citep[for reviews, see][]{2017PhRvD..95l3535C, 2019arXiv190910991L}. Working in a Bayesian framework, as most cosmological analyses do, arguably the most natural way to quantify tension is using the Bayes Ratio~\citep{2006PhRvD..73f7302M}, defined as the ratio of the probability that the two datasets are described by a single set of parameters, to the probability that they are described by separate sets of parameters
\begin{equation}
\label{eq:bayesr}
R = \frac{P(A, B)}{P(A) P(B)} = \frac{\mathcal{Z}_{AB}}{\mathcal{Z}_A \mathcal{Z}_B},
\end{equation}
where $P$ represent a probability, we have omitted the dependence of both probabilities on an underlying model, such as $\Lambda$CDM, and $\mathcal{Z}$ is the Bayesian Evidence. Furthermore, we have assumed that both data sets are independent, an assumption that we further comment on later. High values of $R$ correspond to concordance, and low values are indicative of discordance, with $R$ often interpreted on a Jeffreys' scale~\citep{jeffreys1939theory, 2018PhRvD..98d3526A}.  The main issue of this tension metric, in particular for the analysis of cosmological data sets, is that it is easily proven that $R$ is proportional to the prior volume of shared parameters. Therefore, $R$ cannot be used for analyses that use deliberately flat and wide uninformative priors, such as the analyses of \textit{Planck}, Dark Energy Survey~\citep[DES,][]{2018PhRvD..98d3526A}, Kilo Degree Survey~\citep[KiDS,][]{2020A&A...633A..69H}, ACT, SPT, etc.\ without the arbitrary width of this prior affecting tension assessment. A more detailed interpretation of this discussion can be found in H19.

Motivated by this, H19 defined a new statistic, the {\it Suspiciousness} which keeps all the desired properties of~\cref{eq:bayesr}, but corrects for this undesired dependence on the prior volume. To do so, we divide the Bayes Ratio
in two components: Information and Suspiciousness. The Information is defined as: 
\begin{equation}
\label{eq:i}
\log I = \mathcal{D}_A + \mathcal{D}_B - \mathcal{D}_{AB},
\end{equation}
where $\mathcal{D}$ is the Kullback-Leibler divergence~\cite{KL}. The Information contains the
dependence on the prior volume, therefore by removing it, we obtain a statistic that does not depend on it, 
but is composed of well-defined Bayesian and information theoretic quantities and is therefore covariantly insensitive to reparameterisation of the space. 
Therefore, we define the Suspiciousness as: 
\begin{equation}
    S = \frac{R}{I}.
\end{equation}
In the language of priors, the Suspiciousness may be interpreted as the most cautious Bayes Ratio $R$ corresponding to the narrowest possible priors that do not significantly alter the shape of the posteriors~\citep{2020MNRAS.496.4647L}.

    A significant innovation to the field which we highlight here, first noted in the appendix F.3 of~\citep{2021A&A...646A.140H} and explored in detail in \citep{2021arXiv210211511H} is that since ${\log \mathcal{Z} = \langle \log \mathcal{L} \rangle_\mathcal{P} - \mathcal{D}}$, the suspiciousness can be computed from MCMC chains via
\begin{equation}
    \log S = \langle \log \mathcal{L}_{AB} \rangle_{\mathcal{P}_{AB}} - \langle \log \mathcal{L}_A \rangle_{\mathcal{P}_{A}} - \langle \log \mathcal{L}_B \rangle_{\mathcal{P}_{B}}.
    \label{eqn:chain_def}
\end{equation}
This observation means that so long as one has posterior samples for each of the datasets run separately and in combination, one may compute the suspiciousness without explicitly computing the Bayesian evidence. However, it should be noted that in non-CMB applications only a portion of the parameters are constrained, resulting in hypersurface-like posteriors which are extremely challenging for traditional posterior samplers, but present little challenge for nested samplers.

If the posteriors are such that we may approximate them in the cosmological parameters by a Gaussian (an approximation which is reasonably justified as shown by \cref{fig:parameters}), as derived in H19, if the $d$-dimensional posterior distributions are Gaussian in the parameters with means and covariance $\mu$ and $\Sigma$, then the Suspiciousness is:
\begin{align}
    \log S &= \frac{d}{2}  -\frac{\chi^2}{2}, 
    \label{eqn:logS}\\ 
    \chi^2 &= (\mu_A-\mu_B){(\Sigma_{A}+\Sigma_{B})}^{-1}(\mu_A-\mu_B).
    \label{eqn:chi2}
\end{align}
This may be turned into a tension probability via the survival function of the chi-squared distribution
\begin{equation}
    p =  \int\limits_{\chi^2}^\infty \frac{x^{d/2-1}e^{-x/2}}{2^{d/2}\Gamma(d/2)} \d{x},
    \label{eqn:p}
\end{equation}
and calibrated using a $\sigma$-tension by analogy with the Gaussian case using the inverse of the complementary error function:
\begin{equation}
    \sigma(p) = \sqrt{2}\mathrm{erfc}^{-1}(1-p).
    \label{eqn:sigma}
\end{equation}

Note that, while several methods to quantify tension have been proposed in recent years, they are often built to 
recover \cref{eqn:p} and \cref{eqn:sigma} in the case of Gaussian posterior distributions. Therefore, if this 
work were performed using tension metrics such as Monte-Carlo Parameter Shifts~\cite{2020PhRvD.101j3527R}, Parameter Shifts
in Update Form~\cite{2019PhRvD..99d3506R}, or EigenTension~\cite{2020MNRAS.499.4638P}, we would expect to obtain very similar, 
if not the same results, under the Gaussian approximation used in this work. This is also equivalent to the multivariate measure of tension used in the ACT paper~\cite{2020JCAP...12..047A}.

It should be noted that alternative measures of tension have also been defined and explored that are specialised for the case when two datasets are correlated~\citep{2019MNRAS.484.3126K, 2020PhRvD.101j3527R}. In particular, \cite{2020MNRAS.496.4647L} extended the formalism described in this section to the case of correlated data sets. Applying this to the case of CMB datasets such as \textit{Planck}, ACT, SPT and WMAP (which are correlated by virtue of their measuring the same sky) will form the subject of a future paper.

\vspace{-20pt}
\section{Data}\label{sec:data}

In this work we analyse the three latest CMB data sets, \textit{Planck}, SPT and ACT. As with all cosmological analyses, when considering combining or comparing them at the likelihood level we implicitly assume that the datasets are independent, even though this may not strictly true. Examining the effect of relaxing this assumption will form the subject of future work. It should also be noted that the prior treatment for $\tau$ is different across the three collaborations, one of the aims of future work will be to treat this in a consistent manner for all three cases.
The ACT analysis uses a CMB-derived prior for $\tau$, so there is correlation between posteriors for the $\tau$ parameter. A more complete analysis could adjust the tension in either direction, since correlations in $\tau$ act to reduce the dimensionality to less than $d=6$, increasing the tension~\cite{2019PhRvD.100d3504H,2019PhRvD.100b3512H,2019MNRAS.484.3126K}, but since $\tau$ is a degeneracy-breaking parameter it can have dramatic effects in moving the relative locations of posteriors, increasing or reducing tension.

\subsection{Planck}
\vspace{-10pt}
The \textit{Planck} mission~\cite{2020A&A...641A...1P} was a space observatory that measured the CMB for four years between 2009 and 2013. \textit{Planck} observed the sky in nine frequencies, between 30 and 857 GHz, with the goal of detecting both temperature and polarization anisotropies,
and accurately removing foreground effects. \textit{Planck} measured the power spectrum of temperature anisotropies in multipoles $ \ell \in (2, 2508)$, and for E-mode polarization in multipoles $\ell \in (2, 1996)$, providing the most powerful constraints in the
parameters of the $\Lambda$CDM cosmological model to date.

Beyond the already mentioned tensions in $H_0$ cosmic curvature, and with weak lensing; the most puzzling aspect of the \textit{Planck} analysis
is arguably the $A_{\rm L}$ parameter\footnote{Often known as `lensing parameter' or $A_{\rm lens}$, but we will refrain but these
names as we believe they can be misleading}.
$A_L$ was introduced for internal consistency checks~\citep{2008PhRvD..77l3531C}, and can smooth the peak of the \textit{Planck}
power spectrum. \textit{Planck}~\citep{2020A&A...641A...1P} reports a value $A_L = 1.180 \pm 0.065$ for the combination of temperature and polarization,
meaning that the data seems to prefer more smoothing of the peaks than the best fit $\Lambda$CDM cosmology provides.
While it has been discussed that this could be caused by a statistical fluctuations, especially since the significance 
is lower for different versions of the likelihood~\citep{2019arXiv191000483E}, it has also been hypothesised that it
could be a hint of new physics \cite{2020NatAs...4..196D}, although no theoretical model that produces this effect exists in the literature.
It is important to point out that, while this effect is similar to that of CMB lensing, \textit{Planck} lensing measurements \cite{2020A&A...641A...8P} are compatible with $A_L = 1$. 

Throughout this letter we use the \textit{Planck} legacy archive chains derived using the baseline TTTEEE+low$\ell$+lowE+lensing, and have confirmed that our conclusions are insensitive to excluding the lensing portion of the likelihood.

\vspace{-10pt}
\subsection{South Pole Telescope}
\vspace{-10pt}
We make use of the South Pole Telescope measurements of temperature and polarization from the 500 square degree
analysis of their SPTpol instrument~\cite{2018ApJ...852...97H}. This analysis used data at 150 GHz to produce power spectra for the E-mode polarization (EE) and the temperature-E-mode cross-spectrum (TE).
It should be noted that the TE and EE have been identified as being in disagreement, so strictly this internal combination should also be viewed with suspicion.
The main advantage of SPTpol with respect to \textit{Planck} is its higher resolution, which allows it to measure much smaller scales, 
covering a multipole range $\ell \in (50,8000)$. However, because of its smaller sky coverage, SPTpol cannot obtain information on large scales, and as a consequence produces parameter constraints that are weaker than those from \textit{Planck}. SPT~\citep{2018ApJ...852...97H} reports constraints that differ from \textit{Planck}'s, in particular when only SPTpol's high multipoles are used, but the significance
of this reported discrepancy is not quantified. \textit{Planck}~\cite{2020A&A...641A...1P} used a parameter difference statistic, and 
found no evidence for statistical inconsistencies between the two analyses.  Curiously, performing
an $A_L$ analysis on SPTpol yields a value lower than one, $A_L = 0.81 \pm 0.14$

\vspace{-10pt}
\subsection{Atacama Cosmology Telescope}
\vspace{-10pt}
Finally, we use the Atacama Cosmology Telescope (ACT) posterior samples\footnote{\url{phy-act1.princeton.edu/public/zatkins/ACTPol_lcdm_1.txt}} from Data Release 4 (DR4), which used 6000 square degrees at 98 and 150 GHz 
to produce power spectrum for temperature and polarization extending to $\ell = 4000$. Their results by eye appear to be in 
tension with \textit{Planck}, and ACT~\citep{2020JCAP...12..047A} report a global tension with \textit{Planck} consistent with that recovered in this paper.

\begin{figure*}
	\centerline{%
		\includegraphics{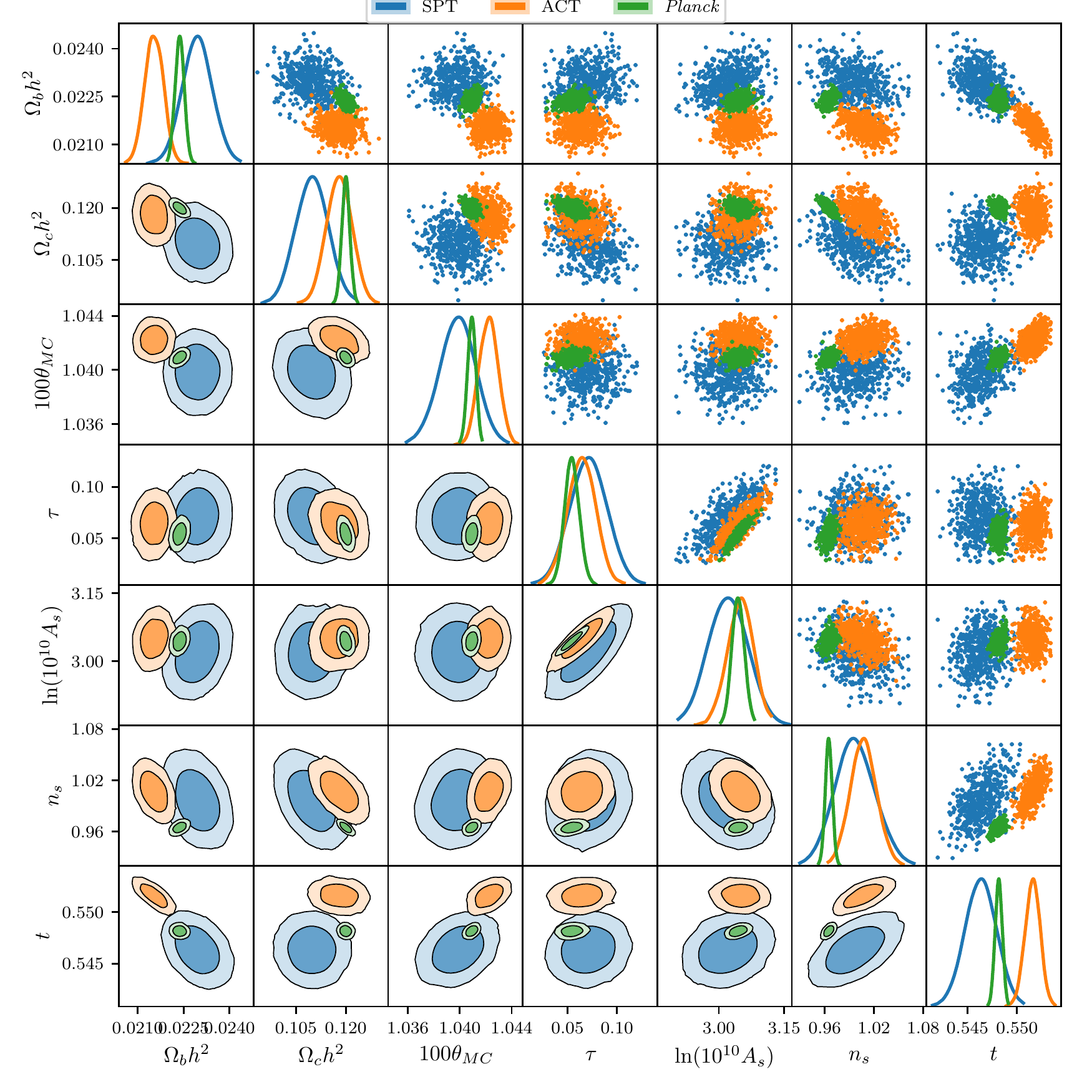}
	}
    \caption{Measurements of the six parameters of the concordance $\Lambda$CDM model using data from the South Pole Telescope (SPT, blue), the Atacama Cosmology Telescope (ACT, orange) and the \textit{Planck} satellite (green). Plots along the diagonal show one-dimensional marginalised probability distributions normalised to equal height, below the diagonal show iso-probability contours containing $68\%$ and $95\%$ of the 2d marginal probability mass, and above the diagonal show samples drawn from the full probability distribution. Of the six cosmological parameters, visually ACT stands out in tension from the other two most clearly in the $n_s-\Omega_bh^2$ plane. We can artificially emphasise this further by computing and plotting the linear combination coordinate of maximum tension between ACT and \textit{Planck} ${t=-\Omega_b h^2 + 0.022 \Omega_c h^2 + 34\theta_{MC} -0.092 \tau + 0.05 {\rm{ln}}(10^{10} A_s) + 0.067 n_s}$, which by construction will have a tension of $\chi=4.15\sigma$. Marginalised plots can therefore over-emphasise tension by ignoring the other active coordinates, but the headline statistics in \cref{tab:tension} are derived from considering the entire distribution as a whole.  Plot produced under \texttt{anesthetic}~\cite{2019JOSS....4.1414H}\label{fig:parameters}}
    \vspace{30pt}
\end{figure*}

\section{Results}\label{sec:results}

Our results are summarised in \cref{tab:tension} and \cref{fig:parameters}. When the Suspiciousness tension quantification techniques are applied to the ACT data products in comparison with the \textit{Planck} baseline, we find a tension probability of ${p=0.86\%}$, with a corresponding Gaussian-calibrated tension of $2.63\sigma$. This level of discrepancy is generally termed mild-to-moderate, and is comparable with some of the larger tensions found between weak lensing and CMB data~\citep[H19,][]{2020A&A...633A..69H}.

The degree of discrepancy between \textit{Planck} and ACT is consistent with the level of tension reported by ACT~\cite{2020JCAP...12..047A}. It is important to note that a global tension quantification such as Suspiciousness does not depend on any specific direction choice in parameter space, nor on the choice of parameters. It also naturally takes into account the effect that in having $d=6$ parameters, it is not improbable that some would be in strong marginal tension by chance. One can make this point explicit by computing an artificial parameter $t$ defined as the linear combination of the other parameters $\theta$ which maximises tension. In the Gaussian case this ``maximum tension parameter'' may be computed as
\begin{equation}
    t \propto {(\mu_A-\mu_B)}^{T} {\left(\Sigma_A + \Sigma_B\right)}^{-1} \theta,
    \label{eqn:tmax}
\end{equation}
and by construction will have a one-dimensional marginalised tension of $\chi$, which in the case of consistency takes the value $t\sim\sqrt{d}\pm 1/\sqrt{2}$. Maximum tension coordinates will be discussed in greater detail in an upcoming work~\cite{liam}.

Marginalised one and two-dimensional projections of the posteriors and the \textit{Planck}-ACT tension coordinate are summarised in \cref{fig:parameters}. It is easy for the eye to be drawn to certain projections where the marginalised tension is large, but as the maximum tension coordinate demonstrates, these can be misleading. We emphasise that the Suspiciousness synthesises all of the posterior information correctly into a single intuitive summary statistic.

Comparing ACT with SPT~\cite{2013PhRvD..88b3501D}, we find a slightly lower mild-to-moderate tension of $2.37\sigma$ ($p=1.8\%$). Interestingly, comparing SPT with \textit{Planck} we find no significant evidence for tension ($p=16.8\%$),  in contradiction with some of the historical literature~\cite{2018ApJ...852...97H}, and in agreement with \cite{2020A&A...641A...1P}. 

Since SPT and \textit{Planck} are consistent, we may confidently combine these datasets. In the absence of a full pipeline run, we combine the Gaussian posterior approximations using Eqs $(14)$--$(20)$ from H19. This \textit{Planck}+SPT combination is $2.79\sigma$ in tension with ACT ($p=0.52\%$), well into the ``moderate'' regime.

Since ACT is in mild-to-moderate tension with both \textit{Planck} and SPT, we should be suspicious of combining it with either, but when we do, as in the final two rows of \cref{tab:tension}, we find no significant evidence for tension, although still higher than when comparing \textit{Planck} and SPT.

In \cref{tab:tension} we also report the $\chi^2$ values for each data combination, and the suspiciousness $\log S$ for reference. As $\log S$ can be regarded as the most conservative value $\log R$ can take by adjusting priors, it is interesting that all values are negative, reflecting the fact that all of the tension probabilities are a little low, when one would traditionally expect $p$ to be uniformly distributed in a frequentist sense, and in general for $d=6$ one would expect positive values of $\log S$ 58\% of the time.

In \cref{tab:true_tension} we compare the full non-Gaussian tension evaluated using \cref{eqn:chain_def} and find the Gaussian approximation to be a slight underestimate of the tension. 

\begin{table}
    \begin{tabular}{ccccc}
        Dataset combination & $\chi^2$ & $p$ & tension & $\log S$ \\
        \hline
        \hline
        ACT vs \textit{Planck}     & $17.2$ & $0.86\%$ & $2.63\sigma$ & $-5.60$ \\  
        ACT vs SPT                 & $15.4$ & $1.77\%$ & $2.37\sigma$ & $-4.68$ \\  
        \textit{Planck} vs SPT     & $9.1$ & $16.82\%$ & $1.38\sigma$ & $-1.55$ \\  
        ACT vs \textit{Planck}+SPT & $18.4$ & $0.52\%$ & $2.79\sigma$ & $-6.22$ \\  
        \hline                                                                 
        ACT+SPT vs \textit{Planck} & $12.2$ & $5.81\%$ & $1.90\sigma$ & $-3.09$ \\  
        ACT+\textit{Planck} vs SPT & $10.3$ & $11.09\%$ & $1.59\sigma$ & $-2.17$ \\ 
    \end{tabular}
    \caption{Global tensions between CMB datasets. For each pairing of datasets we report the $\chi^2$ value calculated using \cref{eqn:chi2}, the corresponding tension probability $p$ from \cref{eqn:p} that such datasets would be this discordant by (Bayesian) chance, a conversion into a Gaussian-equivalent tension using \cref{eqn:sigma} and finall the Suspiciousness from \cref{eqn:logS}. Addition signs in the left column indicate combining the datasets at the likelihood level, and combinations below the line should be viewed with suspicion on account of their discordance reported above the line.\label{tab:tension}}
\end{table}

\begin{table}
    \begin{tabular}{cccc}
        ACT vs \textit{Planck} tension metric & $p$ & tension & $\log S$ \\
        \hline
        True Suspiciousness \cref{eqn:chain_def}       & $0.57\%$ & $2.76\sigma$ & $-6.10$ \\
        Gaussian approximation \cref{eqn:logS}  & $0.86\%$ & $2.63\sigma$ & $-5.60$ \\   
    \end{tabular}
    \caption{We compare the tension computed using the full non-Gaussian expression from \cref{eqn:chain_def}, and the tension computed via the Gaussian approximation. Note that in both cases, for this application since all parameters are well-constrained, all that is required are publicly available MCMC chains.\label{tab:true_tension}}
\end{table}

\vspace{-10pt}
\section{Conclusions}\label{sec:conclusions}
\vspace{-10pt}

In general the causes of tension can be one of three things:
(a) statistical fluctuation 
(b) systematics in at least one of the experiments
(c) evidence for new physics.
Given that we confidently launch manned space missions with higher failure rates than these tensions\footnote{``Bet someone's life'' probabilities can be computed using data from \url{http://www.spacelaunchreport.com/logyear.html}}, as Bayesians we should be very concerned that our CMB measurements are in this much disagreement, so should view statistical fluctuations at this level as a very unsatisfactory explanation. 

The general view (or hope) of many members of the cosmological community at the moment is that the cause of all of these tensions is likely a combination of (b) and (c), and before anyone can claim any kind of new physics we need to get a stronger handle on the systematics in many of our cosmological probes.

As mentioned earlier, this analysis can and will be improved by using a full pipeline of evidences and KL divergences computed using nested sampling~\cite{Skilling}, as well as using techniques that are specialised for dealing with correlated datasets.
However, we would like to draw practitioners' attention in particular to \cref{eqn:chain_def}, which allows them to compute the Suspiciousness using only MCMC chains.

In this letter we do not seek to pass judgement on any of the \textit{Planck}, ACT, or SPT analyses. Indeed, it could be argued that given the quality of all three analyses, it is more likely that these discrepancies indicate a problem with the underlying cosmology, rather than any of the independent pipelines. Combined with the many other tensions emerging between other datasets, the discrepancy quantified in this work lends credence to the possibility that before long we may yet see a paradigm shift in our understanding of the universe.

\section{Acknowledgements}
\begin{acknowledgements}
    WH thanks Gonville \& Caius College for their support via a Research Fellowship. 
	PL thanks STFC \& UCL for their support via a STFC Consolidated Grant.
    Many thanks are accorded to Lukas Hergt for invaluable contributions to the \texttt{anesthetic} package, and to Erminia Calabrese and Daan Meerburg for comments on an early draft. 
\end{acknowledgements}

\bibliographystyle{unsrtnat}
\bibliography{act_planck_tension.bib}

\end{document}